\documentclass[aps,prd,twocolumn,superscriptaddress,preprintnumbers,floatfix,nofootinbib,notitlepage,showkeys,showpacs]{revtex4-1}
\usepackage{datetime}
\usepackage{graphicx}
\usepackage{latexsym}
\usepackage{amsmath,amssymb}
\usepackage{amsfonts}
\newcommand{\SU}{\textrm{SU}\,}

\newcommand{\Tr}{\textrm{Tr}\,}
\newcommand{\ev}[1]{\left \langle #1 \right  \rangle}

\newcommand{\be}{\begin{equation}}
\newcommand{\ee}{\end{equation}}
\newcommand{\bea}{\begin{eqnarray}} % only untightened
\newcommand{\eea}{\end{eqnarray}}

\newcommand{\bmp}{\noindent\begin{minipage}{16cm}}

\begin{document}

\title{Nambu Jona-Lasinio model with Wilson fermions}
\author{Jarno Rantaharju}
\email{rantaharju@cp3-origins.net}
\affiliation{CP3 -Origins \&  IMADA, University of Southern Denmark, Campusvej 55, 5230 Odense, Denmark}
\author{Vincent Drach}
\email{vincent.drach@cern.ch}
\affiliation{CERN, Physics Department, 1211 Geneva 23, Switzerland}
\author{Claudio Pica}
\email{pica@cp3-origins.net}
\affiliation{CP3 -Origins \& IMADA, University of Southern Denmark, Campusvej 55, 5230 Odense, Denmark}
\author{Francesco Sannino}
\email{sannino@cp3-origins.net}
\affiliation{CP3 -Origins \& FKF, University of Southern Denmark, Campusvej 55, 5230 Odense, Denmark}

\begin{abstract}
We present a lattice study of a Nambu Jona-Lasinio (NJL) model using Wilson fermions. Four fermion interactions are a natural part of several extensions of the Standard Model, appearing as a low energy description of a more fundamental theory. In models of dynamical electroweak symmetry breaking they are used to endow the Standard Model fermions with masses. In infrared conformal models these interaction, when sufficiently strong, can alter the dynamics of the fixed point, turning the theory into a (near) conformal model with desirable features for model building. As a first step toward the nonperturbative study of these models, we study the phase space of the ungauged NJL model.
\end{abstract}

\keywords{Lattice Field Theory, The NJL Model, Wilson Fermions}
\pacs{11.15.Ha}

\maketitle

\section{Introduction}

It has been recently shown that gauge Yukawa theories, similar to the Standard Model, even when manifestly perturbative, can abide compositeness conditions \cite{Krog:2015bca}\footnote{This work completes and extends previous results and pioneering work \cite{Bardeen:1989ds,Chivukula:1992pm,Bardeen:1993pj}.}. It was shown that, in certain regions of the gauge Yukawa parameter space, the Higgs-like state is not a propagating degree of freedom at high energies but a low-energy manifestation of an effective four fermion interaction. The theory becomes at some intermediate energies  a gauged version of the celebrated NJL model \cite{Nambu:1961fr}.  

Four fermion interactions emerge naturally in both Technicolor (TC) \cite{Weinberg:1975gm,Susskind:1978ms} and Composite Goldstone Higgs (CH) theories \cite{Kaplan:1983fs,Kaplan:1983sm}  as an effective description at some intermediate energies of a more complete theory of fermion mass generation. A detailed example in which a more fundamental theory, made by only fermions,  yields four fermion interactions that can generate the top mass has been put forward in \cite{Cacciapaglia:2015yra} for the minimal fundamental realization unifying both TC and CH \cite{Cacciapaglia:2014uja}. 
 The four fermion interactions connecting the Higgs sector and the top quark are generally seen to be produced by a high energy gauge or scalar\footnote{Four fermion operators emerge in several other constructions such as  {\it bosonic Technicolor} \cite{Simmons:1988fu,Carone:1992rh,Hemmige:2001vq,Antola:2009wq}  where a TC-singlet elementary Higgs is added to the composite TC-fermion dynamics. 
  One can also naturalize these theories by supersymmetrizing them~\cite{Antola:2010nt}.} 
interactions.  

An alternative popular way to generate masses for the Standard Model fermions   is   known as partial compositeness \cite{Kaplan:1991dc} where each SM fermion $\Psi_{SM}$ couples linearly to a composite fermionic operator ${\cal B}$  through an interaction of the form $\Psi_{SM}\,{\cal B}$. Large anomalous dimensions of the operator ${\cal B}$ (if stemming from purely fermionic fields)  are then invoked such that the operator $ \Psi_{SM}  {\cal B}$ is either  super-renormalizable or marginal. Recent studies of the anomalous dimensions of conformal baryon operators in $\SU(3)$ gauge theories suggest that it is hard to achieve the required anomalous dimensions in purely fermionic theories~\cite{Pica:2016rmv}. Besides the anomalously large anomalous dimensions one needs yet another level of model building to connect the composite baryons to the Standard Model fermions. Reference \cite{Sannino:2016sfx} bypassed these hurdles by constructing a successful example of partial compositeness that makes use of both TC-fermions and TC-scalars. The dynamics of these theories is that that large anomalous dimensions are no-longer needed, one can give masses to all the fermions of the Standard Model, no new model building is required, and therefore they greatly widen the spectrum of theories to investigate on the lattice. If one insists on more involved constructions with only fermions the TC-scalars can be viewed as intermediate composite states. 

It is a fact that whichever is the microscopic extension of the Standard Model it will yield, in certain limits, four fermion interactions  that often reduce to the following three types:
\begin{align*}
L_\text{eff} &= \frac{a}{\Lambda^2_{UV} } (\bar\Psi_{SM}\Psi_{SM})^2  + \frac{b}{\Lambda^2_{UV} } \bar\Psi_{SM}\Psi_{SM}\bar\Psi_{TC} \Psi_{TC}  \\&+ \frac{c}{\Lambda^2_{UV} } (\bar\Psi_{TC}\Psi_{TC})^2.
\end{align*}
The first term, involving only Standard Model fermions, can be suppressed by the cutoff scale $\Lambda_{UV}$, while the other two terms may be enhanced by the dynamics of the technicolor sector.

According to Holdom, \cite{Holdom:1981rm} a model of walking dynamics with a large mass anomalous dimension can enhance the SM fermion mass term dynamically. It was later suggested \cite{Fukano:2010yv} that walking dynamics could be achieved by having the third, Nambu Jona-Lasinio (NJL) type, term induce chiral symmetry breaking in an otherwise infrared conformal Technicolor model \cite{Yamawaki:1996vr,Fukano:2010yv}. We aim therefore to study the nonperturbative dynamics of gauged NJL models. 

We will ultimately study the gauged NJL model with two fermions in the adjoint representation of a SU(2) gauge group. As a first step we investigate an ungauged NJL model on the lattice with Wilson fermions. We only retain the third four fermion term, involving only  TC-fermions.
  A similar model has been studied previously with the goal of understanding the phase structure of Wilson fermions \cite{Bitar:1993cs,Bitar:1993xi,Aoki:1993vs}. Models with staggered fermions have been studied in previous works \cite{Hasenfratz:1991it,AliKhan:1993dx,Kogut:1998rg,Sinclair:1998ji,Catterall:2011ab} and chiral symmetry breaking has been observed. In this study we map the phase space of the model by studying the expectation values of relevant fermion bilinears and the mass spectrum of the lightest mesonic states. The results are qualitatively similar to the meanfield model studied in \cite{Bitar:1993xi}. 
  
This work is a necessary initial step towards a systematic study of four fermion interactions and their impact on models of dynamical symmetry breaking.

\section{The Model}

We study the NJL model with 2 flavors of fermions and 2 colors. The usual action of the NJL model preserves an $SU_L(N_F)\times SU_R(N_F)$ subgroup of the $SU(2N_F)$ flavour symmetry. When representing the fermion fields with pseudofermions, the action must be rendered quadratic using auxiliary fields and the fermion determinant becomes complex\footnote{It is possible to render the fermion determinant positive if the number of colors is even and there is no gauge interaction \cite{Bitar:1993cs,Bitar:1993xi,Aoki:1993vs}. The remedy is not applicable here since we plan to generalize the study to a gauged model. }. We will therefore study a model that preserves just a $U_L(1)\times U_R(1)$ subgroup of the flavor symmetry.

The model with a nonzero quark mass is defined by the Lagrangian
\begin{align}
 \tilde L(x) = &\bar\Psi (x) \left [ D_W + m_0+\sigma(x) + \pi_3(x) i \gamma_5\tau_3 \right ] \Psi(x) \label{cont_tL} \\
 &+\frac{\sigma(x)^2+\pi_3(x)^2}{4\gamma^2}, \nonumber
\end{align}
where $D_W$ is the Wilson Dirac operator.
After integrating out the auxiliary fields we recover the Lagrangian
\begin{align}
 L(x) = &\bar\Psi(x) \left ( D_W +m_0 \right ) \Psi(x) \\ 
 &- \gamma^2 \left [ \left( \bar\Psi(x)\Psi(x) \right )^2 + \left( \bar\Psi(x) i\gamma_5\lambda^3 \Psi(x) \right )^2 \right ] \nonumber
\end{align}
and the equations of motion for the auxiliary fields are
\begin{align}
\ev{\sigma(x)} &= -2\gamma^2 \,\ev{\bar\Psi(x)\Psi(x)}, \\
 \ev{\pi_3(x)} &= -2\gamma^2 \ev{\bar\Psi(x)  i\gamma_5 \tau_3 \Psi(x)}. \label{aux_exps}
\end{align}

\begin{figure}[t] \center
\includegraphics[width=.45\linewidth,height=.455\linewidth]{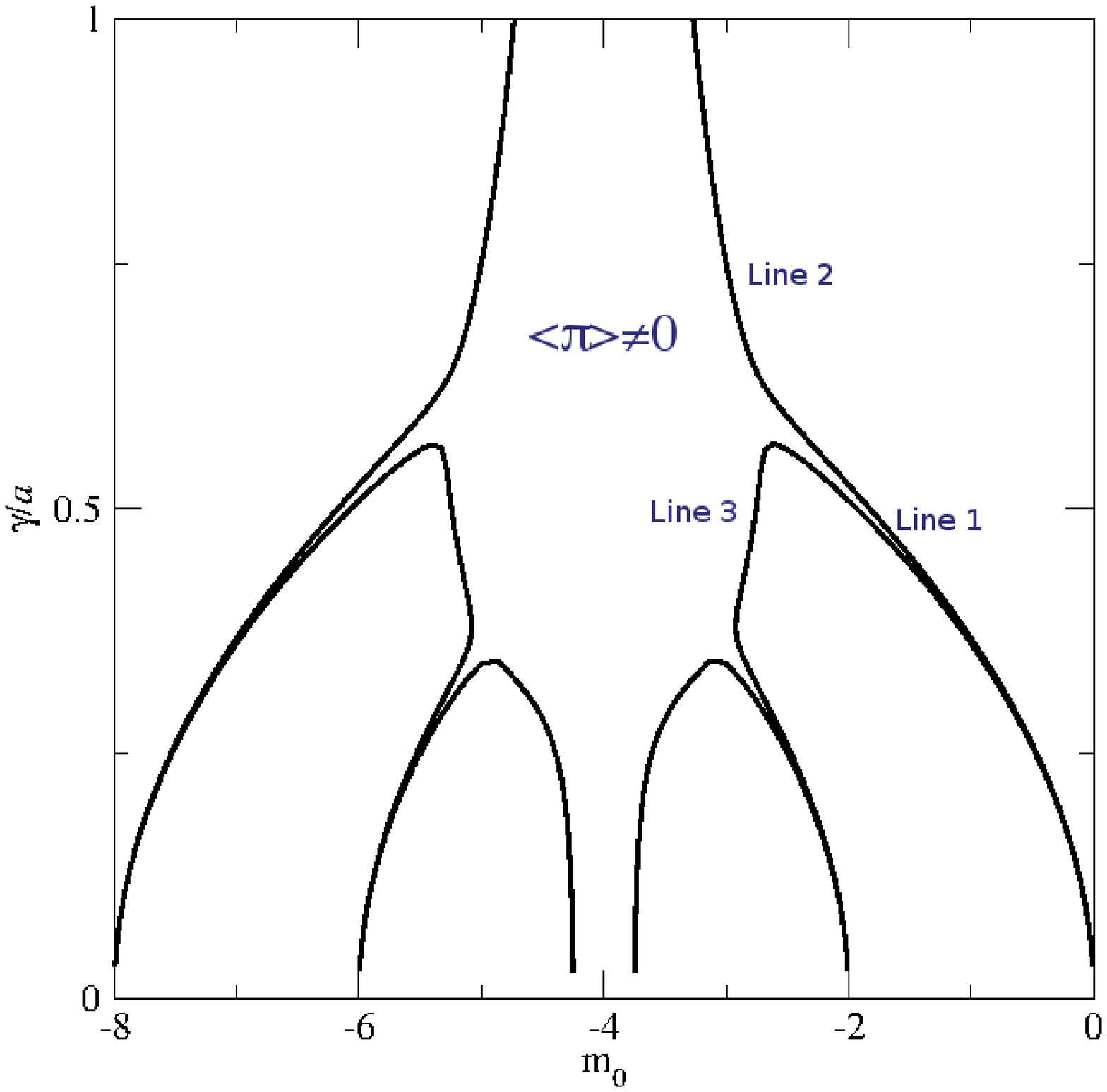}
\includegraphics[width=.45\linewidth,height=.45\linewidth]{{phase_gamma2_zoomed}.eps}
\caption{ The phase diagram of the lattice NJL-model. There is a flavor-parity broken phase around $m_0=-4$, where the expectation value of $\pi_3$  is nonzero. This phase is surrounded by critical lines where the pseudoscalar meson has zero mass. In the zoomed plot on the right, we compare the meanfield estimate to the numerical results described in section \ref{numericalstudies}. The red dashed lines show the range of masses studied at each $\gamma$. The open circles represent the observed critical line in the chirally symmetric phase (line 1). The crosses show an unphysical chirally broken critical line (line 3) and the asterisks show a physical chirally broken critical line (line 2). Finally, at the point marked with both a circle and a cross, the status of spontaneous chiral symmetry breaking is unclear. }
\label{meanfield_phaseplot}
\end{figure}

It is useful to gain insight into the model via  meanfield computations \cite{Bitar:1993cs,Bitar:1993xi,Aoki:1993vs}. A sketch of the phase diagram is shown in the left panel of figure \ref{meanfield_phaseplot} with the lattice size $8^3\times16$. The right panel shows a comparison to the numerical results in the next section. The solid lines in the figure show second order transitions where the auxiliary field $\pi_3$ develops an expectation value. Inside the region outlined by the critical lines, around $m_0=-4$, the expectation value $\ev{\pi_3} \neq 0$ and parity and flavor symmetries are broken. The lines also correspond to a zero pseudoscalar meson mass, and therefore to zero quark mass and the restoration of chiral symmetry.

Line 1 corresponds to the restoration of chiral symmetry in the unbroken phase. The parity broken phase below line 1 is narrow and disappears at the infinite volume limit. There is only a small change in $\ev{\sigma}$ when crossing this phase. Line 2 corresponds to the critical line with spontaneously broken chiral symmetry. The parity broken phase is wider and since the model is symmetric around $m_0=-4$, $\ev{\sigma}$ changes sign across the broken phase. The critical coupling is close to $\gamma=0.55a$.

\section{Numerical Results} \label{numericalstudies}

In order to study the model from first principles, we generate configurations of $\sigma(x)$ and $\pi_3(x)$ using the Hybrid MonteCarlo (HMC) algorithm. We consider lattices of size $V=a^4L^3T$, with $L=8$ and $T=16$, except for a few simulations studying volume scaling. We use a second order integrator with trajectory length $t_{HMC}=1$. The step size is selected so that the acceptance rate is above $0.8$.

\begin{figure}[t]
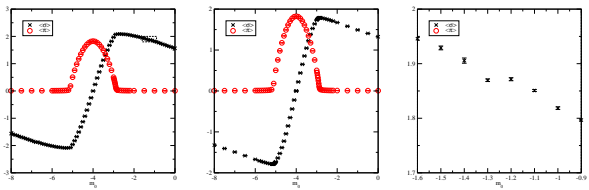
 \center
{\includegraphics[width=.3\linewidth]{{sigma_pi_s0.8}.eps}}
{\includegraphics[width=.3\linewidth]{{sigma_pi_s1.2}.eps}}
{\includegraphics[width=.3\linewidth]{{sigma_pi_s0.8_zoom}.eps}}
\caption{ The expectation values of the auxiliary fields with varying $m_0$ and $\gamma=0.4a$ (left) and $\gamma=0.6a$ (middle). The plot of right shows the region marked by dashed lines in the leftmost one.  The parity broken phase is clearly marked by the nonzero expectation value of the $\pi_3$ field. The model is symmetric around $m_0=-4$ and we see the see the condensate $\ev{\sigma}$ change sign when crossing the parity broken phase. In the zoomed plot on the right we see a small discontinuity in $\ev{\sigma}$, which is expected on a finite lattice when chiral symmetry is not broken.}
\label{pisigmas}
\end{figure}

First we study the phase diagram by measuring the volume averaged expectation values of the $\pi_3$ and and $\sigma$ fields. The results shown have been obtained from 100 HMC trajectories after thermalization. We choose 6 values of $\gamma$ from $0.4a$ to $0.65a$.  In figure \ref{pisigmas} we show their behavior at two representative values of the coupling, $\gamma=0.4a$, which lies on the chirally symmetric side, and $\gamma=0.6a$, which is on the broken side. The critical lines observed are also shown in the right panel in figure \ref{meanfield_phaseplot}.

The expectation value 
\begin{align}
\ev{\pi} = \frac1V \ev{ \left | \sum_x \pi_3(x) \right | }
\end{align}
indeed becomes nonzero on lines 2 and 3. Line 1, however, is not observed from the behavior of $\ev{\pi}$. The expectation value is likely to be too small, or the broken phase too narrow, to be observed with the current precision. This critical line can be identified by studying the expectation value
\begin{align}
\ev{\sigma} = \frac1V\ev{ \sum_x \sigma(x)}.
\end{align}
 This quantity is related to the chiral condensate and has a discontinuity one the line 1 if the boundary conditions for the fermion fields are periodic. The discontinuity observed at $\gamma=0.4a$ is shown in figure \ref{pisigmas} in the third panel. The measurable $\ev{\sigma}$ also  changes behavior on the other critical lines, decreasing as $\ev{\pi}$ increases and changing sign at $m_0=-4$.

\begin{figure}[t]
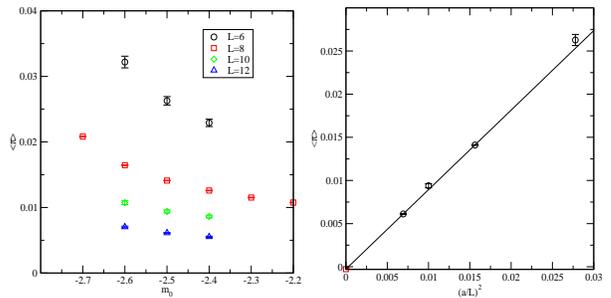
 \center
\includegraphics[width=.45\linewidth,height=.45\linewidth]
{{pi_L_symm}.eps}
\includegraphics[width=.45\linewidth,height=.45\linewidth]
{{pi_L_scaling}.eps}
\caption{ The order parameter $\ev{\pi}$ with $\gamma=0.6a$ on the symmetric (positive mass) side of line 2 (left) and a second order infinite volume extrapolation at $m_0=-2.5$ (right).   }
\label{pi_with_L}
\end{figure}

\begin{figure}[t]
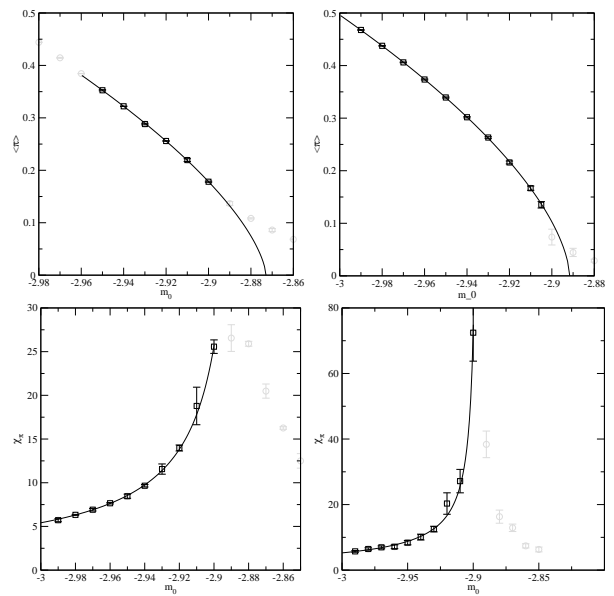
 \center
\includegraphics[width=.45\linewidth,height=.45\linewidth]{{pi_dimension_L8}.eps}
\includegraphics[width=.45\linewidth,height=.45\linewidth]{{pi_dimension_L12}.eps}
\includegraphics[width=.45\linewidth,height=.45\linewidth]{{pi_susc_left_dimension_L8}.eps}
\includegraphics[width=.45\linewidth,height=.45\linewidth]{{pi_susc_left_dimension_L12}.eps}
\caption{ The order parameter $\ev{\pi}$ and the susceptibility $\chi_\pi$ at the transition on line 2 with $\gamma=0.6a$ and $L=8$ (left) and $L=12$ (right).  }
\label{pi_scaling}
\end{figure}

The order of the transition on the line 2 is of special interest. If the transition is second order the correlation length of the order parameter $\ev{\pi_3} = 2\gamma^2\ev{\bar\psi \gamma_5i\tau_3\psi}$ diverges and thus the mass of the corresponding pseudoscalar meson is zero. We have verified that the order parameter is zero on the symmetric side by measuring it at a few lattice sizes and studying the scaling. The values are shown in figure \ref{pi_with_L}. We have then studied the value of the order parameter in the critical region with $L=8$ and $L=12$. Scaling fits to the function
\begin{align}
 \ev{\pi} = C_\pi \left | m_0-m_c  \right |^{\beta}
\end{align}
are shown in figure \ref{pi_scaling}. We find $\beta=0.65(2)$ and $0.56(2)$ with $L=8$ and $12$ respectively. The susceptibility
\begin{align}
\chi_\pi = V \left ( \ev{\pi^2} - \ev{\left | \pi \right | }^2 \right )
\end{align}
is strongly peaked at the transition and performing a similar fit to
\begin{align}
\chi_\pi = C_\chi \left | m_0 - m_c \right |^{-\nu} 
\end{align}
we find $\nu=0.83(5)$ and $\nu=0.90(6)$ with $L=8$ and $12$ respectively.
 
The lattice value for the chiral condensate is related to $\ev{\sigma}$ by equation \ref{aux_exps}. As usual, the chiral condensate suffers from additive renormalization and the renormalized value cannot be read directly from the measured value of $\ev{\sigma}$. Nevertheless the large change in the behaviour of $\ev{\sigma}$ when crossing line 2, as compared to the small discontinuity on line 1, points to a first order transition between a chirally symmetric and a broken phase. This can be verified in a straightforward way by studying the mass of the Goldstone boson of the symmetry breaking and comparing to other meson masses.

Since the spontaneous breaking of the chiral symmetry breaks only one generator, there is only one Goldstone boson. This is a pseudoscalar meson related to the diagonal subgroup of the isospin triplet. There are four additional pseudoscalar mesons, two of which are also components of the isospin triplet. The difference between the diagonal meson and the others is encoded in a disconnected contribution in the channel, directly related to the field $\pi_3(x)$.  More details on the evaluation of the disconnected contribution is given in appendix \ref{appendix_disconnected}. To reduce the noise in the disconnected channel of the pseudoscalar case we measure the correlator using two interpolating operators with the generalized eigenvalue method and use hopping parameter expansion in the inversion of the fermion matrix. In the case of the vector meson, the disconnected contribution does not present a problem.

For each parameter set we generate between 2000 and 20000 configurations separated by 20 HMC trajectories. Even with the large number of measurements we must note that in many cases we do not reach a plateau in the effective mass and that there may be systematic errors in the pseudoscalar masses larger than the statistical errors. The error is less than 10\% and we consider our accuracy sufficient for an exploratory study of the phase diagram.

The field $\pi_3({\bf x},t)$ can also serve as an interpolating operator for the pseudoscalar meson. The evaluation of this correlation function does not require inverting the fermion matrix and is therefore efficient. We measure it using between 20000 and 100000 configurations for each parameter set separated by 10 HMC trajectories. The result is noisy at large mass and at small $\gamma$. We report this measurement as $m_{\pi_2}$ when it can be performed with sufficient accuracy.

It is worth noting that the diagonal pseudoscalar meson mass is not necessary for studying the phase diagram.
The disconnected contribution is small in the vector correlation function  and absent in the nondiagonal triplet channels. The masses of these mesons can be estimated accurately with substantially less data. The critical line on the chirally symmetric side (line 1) can be identified easily by finding the bare mass where all masses are zero. On the broken side these masses should remain nonzero at the critical line (line 2). Here the critical line can be identified from the expectation value $\ev{\pi}$ and as long as the transition is second order, the pseudoscalar mass is zero on the critical line.

\begin{figure}[t] \center
\includegraphics[width=.45\linewidth,height=.45\linewidth]{{mpirho0.8}.eps}
\includegraphics[width=.45\linewidth,height=.45\linewidth]{{mpirho0.9}.eps}
\caption{ The pseudoscalar and vector meson masses with the coupling $\gamma=0.4a$ (left) and $0.45a$ (right). We also show the expectation value $\ev{\pi}$, which serves as the order parameter for the parity broken phase.}
\label{large_coupling_masses1}

\vspace*{\floatsep}

\includegraphics[width=.45\linewidth,height=.45\linewidth]{{mpirho1}.eps}
\includegraphics[width=.45\linewidth,height=.45\linewidth]{{mpirho1.1}.eps}
\caption{ The pseudoscalar and vector meson masses and $\ev{\pi}$ with the coupling $\gamma=0.5a$ (left) and $0.55a$ (right).}
\label{large_coupling_masses2}

\vspace*{\floatsep}

\includegraphics[width=.45\linewidth,height=.45\linewidth]{{mpirho1.2}.eps}
\includegraphics[width=.45\linewidth,height=.45\linewidth]{{mpirho1.3}.eps}
\caption{ The pseudoscalar and vector meson masses and $\ev{\pi}$ with the coupling $\gamma=0.6a$ (left) and $0.65a$ (right).}
\label{large_coupling_masses3}
\end{figure}

In figures \ref{large_coupling_masses1}, \ref{large_coupling_masses2} and \ref{large_coupling_masses3} we show the vector meson mass ($m_\rho$) and diagonal pseudoscalar mass measured from the usual fermionic correlator ($m_\pi$) and from the correlator of the field $\pi_3$ ($m_{\pi_2}$). We also show the order parameter for the parity broken phase $\ev{\pi_3}$. We study finite size effects by measuring the masses with lattice size $24\times12^3$ at a few interesting values of $m_0$ and $\gamma$.
At small coupling, $0.4a \leq \gamma \leq 0.5a$, we see the two expected critical lines. At large $m_0$ the pseudoscalar and the vector masses are identical and approach zero linearly around the first critical line, line 1 in figure \ref{meanfield_phaseplot}. On the negative mass side, the vector and pseudoscalar masses split and the pseudoscalar mass becomes zero at a second critical line, line 3 in figure \ref{meanfield_phaseplot}. At the second critical line the model enters the wider parity broken region and we see a nonzero value for $\ev{\pi}$. The difference between $m_\rho$ and $m_\pi$ at the line 3 implies a nonzero chiral condensate.

At $\gamma=0.55a$ the two critical lines have merged. At current accuracy it is not possible to tell chiral symmetry is broken in this case. The results from the larger lattice size, however, show no splitting between the vector and pseudoscalar masses. At large coupling, $\gamma=0.6a$ and $0.65a$, we observe only one critical line, corresponding to line 2 in figure \ref{meanfield_phaseplot}. At the critical line, the vector mass remains nonzero, while the pseudoscalar mass becomes zero, implying that a condensate has formed and the chiral symmetry is broken.

\section{Conclusions}

We have studied the phase space of the Nambu Jona-Lasinio model with Wilson fermions. By measuring the expectation values of the auxiliary fields $\ev{\sigma}$ and $\ev{\pi}$, related respectively to the chiral condensate and a flavor-parity breaking condensate, we were able to identify the critical lines with zero quark mass.

We have measured the masses of the Goldstone boson when chiral symmetry breaks as well the vector meson as  functions of the bare mass and coupling. We find evidence of the formation of a condensate and chiral symmetry breaking above the critical coupling $\gamma\approx 0.55a$. In the chirally symmetric phase the vector meson mass is zero at the critical line and provides a convenient way of identifying it. In the chirally broken phase the vector meson mass remains nonzero, but the critical line can be identified by the formation of the parity breaking condensate $\ev{\pi}$. The Goldstone boson mass is always zero on the critical lines, but it is inconvenient to measure due to disconnected contributions.

The phase structure of the model, as a function of $m_0$ and $\gamma$ is shown in figure \ref{meanfield_phaseplot}, along with the meanfield estimate \cite{Bitar:1993cs,Bitar:1993xi,Aoki:1993vs}. The critical lines are indicated by local minima in the pseudoscalar meson mass as a function of $m_0$. The breaking of chiral symmetry is indicated by a nonzero value of the vector meson mass at the critical line.

The result agrees qualitatively with the meanfield estimate. The exact location of the critical line deviates noticeably from the estimate, with the difference increasing with the coupling. In general the effect is that of increasing $\gamma$.

Since the phase structure of the model can be mapped clearly and agrees well with expectation, we conclude that any systematic effects are under control. We have observed chiral symmetry breaking in the model and can distinguish between the chirally symmetric and broken phases. The next step is to include the gauge interaction and study the phase diagram the gauged NJL model with 2 fermion is the adjoint representation of a SU(2) gauge group.

\section{Acknowledgments}

This work was supported by the Danish National Research Foundation DNRF:90 grant and by a Lundbeck Foundation Fellowship grant. The computing facilities were provided by the Danish Centre for Scientific Computing and the DeIC national HPC center at SDU.

\appendix 
\section{ Disconnected Diagrams }
\label{appendix_disconnected}

The Goldstone boson of spontaneous chiral symmetry breaking, the diagonal isotriplet pseudoscalar meson, differs from other pseudoscalar mesons by a disconnected term in the propagator.  The term is directly related to the auxiliary field $\pi_3$ and disappears with zero four fermion coupling. Disconnected contributions arise also in other diagonal isotriplet channels, but appear to be negligible in the vector channel.

The isotriplet meson masses are measured from correlators of the type
\begin{align}
&C_\Gamma(t_0) \label{discon_props} \\
&= \frac{1}{V_3} \ev{ \sum_{y}\left(\bar\Psi(0,0)\Gamma \tau_a \Psi(0,0)\right)^\dagger \bar\Psi(y,t_0)\Gamma \tau_a \Psi(y,t_0) } \nonumber \\
&= -\frac{1}{V_3} \ev{ \sum_{y} \Tr\left[ \left(S(y,t_0;0,0) \Gamma \tau_a \right )^\dagger S(0,0;y,t_0) \Gamma \tau_a  \right ] }  \nonumber \\
&+ \frac{1}{V_3} \ev{ \sum_{x,y,t} \Tr\left[ S(0,0;0,0) \Gamma \tau_a \right]^\dagger \Tr\left[  S(y,t_0;y,t_0) \Gamma \tau_a  \right ] }, \nonumber
\end{align}
where $\tau_a$ are Pauli matrices in flavour space. The second term on the right hand side in equation \ref{discon_props} is called the disconnected contribution. The propagator $S$ is diagonal in flavour space and the trace is clearly zero when $a=1,2$.
With $a=3$ the trace becomes $\Tr  [S\tau_3]=\Tr [S_u-S_d]$. This can be nonzero when the pseudoscalar auxiliary field $\pi_3$ is nonzero.

Writing the propagator as 
\begin{align}
S_{u,d}&= \frac{1}{M_{u,d}} = \frac{M^\dagger_{u,d}}{M_{u,d}^\dagger M_{u,d}}\\
&= \frac{\sum_\mu \partial_{\mu} \gamma_\mu + \left ( \sigma \pm i\pi_3 \gamma_5 \right)}{M^\dagger M}
\end{align}
the disconnected part is
\begin{align}
\Tr \left(S_u-S_d\right)\Gamma &= \Tr \frac{2 i\pi_3\gamma_5 }{M^\dagger M}\Gamma.
\end{align}
At $\gamma=0$ the auxiliary field $\pi_3$ is restricted to zero and the disconnected contribution disappears.

\begin{figure}[h]
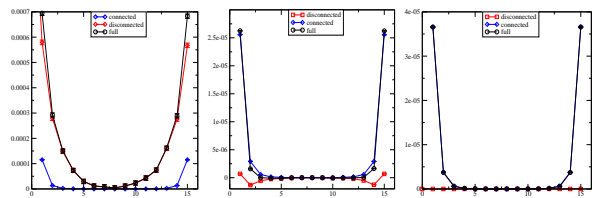
 \center
\includegraphics[width=.29\linewidth,height=.29\linewidth]{{m2.9_g5_prop}.eps}
\includegraphics[width=.29\linewidth,height=.29\linewidth]{{m2.9_g0g5_prop}.eps}
\includegraphics[width=.29\linewidth,height=.29\linewidth]{{m2.9_g1_prop}.eps}
\caption{
The connected and disconnected contributions to the isotriplet channels with $\Gamma=\gamma_5$, $\Gamma=\gamma_0 \gamma_5$ and the $\Gamma=\gamma_k$ from left to right with $\gamma=0.65a$ and $m_0=-2.9$. The two pseudoscalar channels show significant disconnected contributions. The contributions have different weights in the different channels and we use the generalized eigenvalue method to differentiate the contributions. The disconnected contribution in the vector channel is zero at current accuracy.} \label{disconnected}
\end{figure}

We observe that the disconnected term is not significant in the vector channel. This may be understood in perturbation theory around $\gamma=0$: the disconnected term appears only at fourth order in $\gamma$ in the vector channel, but arises at the second order in the pseudoscalar channel. In figure \ref{disconnected} we compare the connected and disconnected contributions in the pseudoscalar and vector channels at $\gamma =  0.65a$, $L=8^3\times16$ and $m_0=2.9$. 

While the disconnected term appears in all diagonal isotriplet correlators, it has a different weight in different channels. The $\Gamma=\gamma_5$ channel mixes maximally with the $\pi_3$ field and has a large disconnected contribution. In the $\Gamma=\gamma_0\gamma_5$ channel the contribution is somewhat smaller. In most cases we use the generalized eigenvalue method with in the space of these two channels to measure the pseudoscalar meson mass. At small coupling $\gamma=0.4a$ and $0.45a$ and $am_0>-2.4$, the field $\pi_3$ has large variation and the first channel becomes noisy. In this region we use only the second channel.


\begin{thebibliography}{99}
 

%\cite{Foadi:2012bb}
%\bibitem{Foadi:2012bb}
%  R.~Foadi, M.~T.~Frandsen and F.~Sannino,
  %``125 GeV Higgs boson from a not so light technicolour scalar,''
%  Phys.\ Rev.\ D {\bf 87} (2013) 9,  095001
%  [arXiv:1211.1083 [hep-ph]].
  %%CITATION = ARXIV:1211.1083;%%
  %57 citations counted in INSPIRE as of 28 Aug 2015
  
  %\cite{DiChiara:2014gsa}
%\bibitem{DiChiara:2014gsa}
%  S.~Di Chiara, R.~Foadi and K.~Tuominen,
  %``125 GeV Higgs from a chiral techniquark model,''
%  Phys.\ Rev.\ D {\bf 90} (2014) 11,  115016
%  [arXiv:1405.7154 [hep-ph]].
  %%CITATION = ARXIV:1405.7154;%%
  %4 citations counted in INSPIRE as of 28 Aug 2015
  
  %\cite{Krog:2015bca}
\bibitem{Krog:2015bca} 
  J.~Krog, M.~Mojaza and F.~Sannino,
  %``Four-Fermion Limit of Gauge-Yukawa Theories,''
  Phys.\ Rev.\ D {\bf 92}, no. 8, 085043 (2015)
  doi:10.1103/PhysRevD.92.085043
  [arXiv:1506.02642 [hep-ph]].
  %%CITATION = doi:10.1103/PhysRevD.92.085043;%%
  %7 citations counted in INSPIRE as of 11 Jul 2016
  
  
  %\cite{Bardeen:1989ds}
\bibitem{Bardeen:1989ds} 
  W.~A.~Bardeen, C.~T.~Hill and M.~Lindner,
  %``Minimal Dynamical Symmetry Breaking of the Standard Model,''
  Phys.\ Rev.\ D {\bf 41}, 1647 (1990).
  doi:10.1103/PhysRevD.41.1647
  %%CITATION = doi:10.1103/PhysRevD.41.1647;%%
  %1088 citations counted in INSPIRE as of 11 Jul 2016
   
%  
%  %\cite{Gerhold:2007yb}
%\bibitem{Gerhold:2007yb} 
%  P.~Gerhold and K.~Jansen,
%  %``The Phase structure of a chirally invariant lattice Higgs-Yukawa model for small and for large values of the Yukawa coupling constant,''
%  JHEP {\bf 0709}, 041 (2007)
%  doi:10.1088/1126-6708/2007/09/041
%  [arXiv:0705.2539 [hep-lat]].
%  %%CITATION = doi:10.1088/1126-6708/2007/09/041;%%
%  %39 citations counted in INSPIRE as of 11 Jul 2016
%  
%  
  %\cite{Chivukula:1992pm}
\bibitem{Chivukula:1992pm} 
  R.~S.~Chivukula, M.~Golden and E.~H.~Simmons,
  %``Critical constraints on chiral hierarchies,''
  Phys.\ Rev.\ Lett.\  {\bf 70}, 1587 (1993)
  doi:10.1103/PhysRevLett.70.1587
  [hep-ph/9210276].
  %%CITATION = doi:10.1103/PhysRevLett.70.1587;%%
  %28 citations counted in INSPIRE as of 11 Jul 2016
  
  %\cite{Bardeen:1993pj}
\bibitem{Bardeen:1993pj} 
  W.~A.~Bardeen, C.~T.~Hill and D.~U.~Jungnickel,
  %``Chiral Hierarchies, Compositeness and the Renormalization Group,''
  Phys.\ Rev.\ D {\bf 49}, 1437 (1994)
  doi:10.1103/PhysRevD.49.1437
  [hep-th/9307193].
  %%CITATION = doi:10.1103/PhysRevD.49.1437;%%
  %26 citations counted in INSPIRE as of 11 Jul 2016
  
  %\cite{Nambu:1961fr}
\bibitem{Nambu:1961fr} 
  Y.~Nambu and G.~Jona-Lasinio,
  %``Dynamical Model Of Elementary Particles Based On An Analogy With Superconductivity. Ii,''
  Phys.\ Rev.\  {\bf 124}, 246 (1961).
  doi:10.1103/PhysRev.124.246
  %%CITATION = doi:10.1103/PhysRev.124.246;%%
  %2245 citations counted in INSPIRE as of 11 Jul 2016
  
  %\cite{Weinberg:1975gm}
\bibitem{Weinberg:1975gm} 
  S.~Weinberg,
  %``Implications of Dynamical Symmetry Breaking,''
  Phys.\ Rev.\ D {\bf 13}, 974 (1976).
  doi:10.1103/PhysRevD.13.974
  %%CITATION = doi:10.1103/PhysRevD.13.974;%%
  %1471 citations counted in INSPIRE as of 11 Jul 2016
  
  %\cite{Susskind:1978ms}
\bibitem{Susskind:1978ms} 
  L.~Susskind,
  %``Dynamics of Spontaneous Symmetry Breaking in the Weinberg-Salam Theory,''
  Phys.\ Rev.\ D {\bf 20}, 2619 (1979).
  doi:10.1103/PhysRevD.20.2619
  %%CITATION = doi:10.1103/PhysRevD.20.2619;%%
  %2338 citations counted in INSPIRE as of 11 Jul 2016
  
  %\cite{Kaplan:1983fs}
\bibitem{Kaplan:1983fs} 
  D.~B.~Kaplan and H.~Georgi,
  %``SU(2) x U(1) Breaking by Vacuum Misalignment,''
  Phys.\ Lett.\ B {\bf 136}, 183 (1984).
  doi:10.1016/0370-2693(84)91177-8
  %%CITATION = doi:10.1016/0370-2693(84)91177-8;%%
  %544 citations counted in INSPIRE as of 11 Jul 2016
  
  %\cite{Kaplan:1983sm}
\bibitem{Kaplan:1983sm} 
  D.~B.~Kaplan, H.~Georgi and S.~Dimopoulos,
  %``Composite Higgs Scalars,''
  Phys.\ Lett.\ B {\bf 136}, 187 (1984).
  doi:10.1016/0370-2693(84)91178-X
  %%CITATION = doi:10.1016/0370-2693(84)91178-X;%%
  %436 citations counted in INSPIRE as of 11 Jul 2016
  
  
%\cite{Cacciapaglia:2015yra}
\bibitem{Cacciapaglia:2015yra} 
  G.~Cacciapaglia and F.~Sannino,
  %``An Ultraviolet Chiral Theory of the Top for the Fundamental Composite (Goldstone) Higgs,''
  Phys.\ Lett.\ B {\bf 755}, 328 (2016)
  doi:10.1016/j.physletb.2016.02.034
  [arXiv:1508.00016 [hep-ph]].
  %%CITATION = doi:10.1016/j.physletb.2016.02.034;%%
 
 
 %\cite{Cacciapaglia:2014uja}
\bibitem{Cacciapaglia:2014uja} 
  G.~Cacciapaglia and F.~Sannino,
  %``Fundamental Composite (Goldstone) Higgs Dynamics,''
  JHEP {\bf 1404}, 111 (2014)
  doi:10.1007/JHEP04(2014)111
  [arXiv:1402.0233 [hep-ph]].
  %%CITATION = doi:10.1007/JHEP04(2014)111;%%
  %47 citations counted in INSPIRE as of 11 Jul 2016
 
 
 %\bibitem{Foadi:2012bb}
%  R.~Foadi, M.~T.~Frandsen and F.~Sannino,
  %``125 GeV Higgs boson from a not so light technicolour scalar,''
%  Phys.\ Rev.\ D {\bf 87} (2013) 9,  095001
%  [arXiv:1211.1083 [hep-ph]].
  %%CITATION = ARXIV:1211.1083;%%
  %57 citations counted in INSPIRE as of 28 Aug 2015
  
  %\cite{DiChiara:2014gsa}
%\bibitem{DiChiara:2014gsa}
%  S.~Di Chiara, R.~Foadi and K.~Tuominen,
  %``125 GeV Higgs from a chiral techniquark model,''
%  Phys.\ Rev.\ D {\bf 90} (2014) 11,  115016
%  [arXiv:1405.7154 [hep-ph]].
  %%CITATION = ARXIV:1405.7154;%%
  %4 citations counted in INSPIRE as of 28 Aug 2015
  
  
  %\cite{Simmons:1988fu}
\bibitem{Simmons:1988fu} 
  E.~H.~Simmons,
  %``Phenomenology of a Technicolor Model With Heavy Scalar Doublet,''
  Nucl.\ Phys.\ B {\bf 312}, 253 (1989).
  doi:10.1016/0550-3213(89)90296-4
  %%CITATION = doi:10.1016/0550-3213(89)90296-4;%%
  %96 citations counted in INSPIRE as of 11 Jul 2016
  
     
  
  %\cite{Carone:1992rh}
\bibitem{Carone:1992rh} 
  C.~D.~Carone and E.~H.~Simmons,
  %``Oblique corrections in technicolor with a scalar,''
  Nucl.\ Phys.\ B {\bf 397}, 591 (1993)
  doi:10.1016/0550-3213(93)90187-T
  [hep-ph/9207273].
  %%CITATION = doi:10.1016/0550-3213(93)90187-T;%%
  %59 citations counted in INSPIRE as of 11 Jul 2016
  
  %\cite{Hemmige:2001vq}
\bibitem{Hemmige:2001vq} 
  V.~Hemmige and E.~H.~Simmons,
  %``Current bounds on technicolor with scalars,''
  Phys.\ Lett.\ B {\bf 518}, 72 (2001)
  doi:10.1016/S0370-2693(01)01031-0
  [hep-ph/0107117].
  %%CITATION = doi:10.1016/S0370-2693(01)01031-0;%%
  %15 citations counted in INSPIRE as of 11 Jul 2016
  
  
  %\cite{Antola:2009wq}
\bibitem{Antola:2009wq} 
  M.~Antola, M.~Heikinheimo, F.~Sannino and K.~Tuominen,
  %``Unnatural Origin of Fermion Masses for Technicolor,''
  JHEP {\bf 1003}, 050 (2010)
  doi:10.1007/JHEP03(2010)050
  [arXiv:0910.3681 [hep-ph]].
  %%CITATION = doi:10.1007/JHEP03(2010)050;%%
  %34 citations counted in INSPIRE as of 11 Jul 2016
  
  %\cite{Antola:2010nt}
\bibitem{Antola:2010nt} 
  M.~Antola, S.~Di Chiara, F.~Sannino and K.~Tuominen,
  %``Minimal Super Technicolor,''
  Eur.\ Phys.\ J.\ C {\bf 71}, 1784 (2011)
  doi:10.1140/epjc/s10052-011-1784-1
  [arXiv:1001.2040 [hep-ph]].
  %%CITATION = doi:10.1140/epjc/s10052-011-1784-1;%%
  %24 citations counted in INSPIRE as of 11 Jul 2016
  
  %\cite{Kaplan:1991dc}
\bibitem{Kaplan:1991dc} 
  D.~B.~Kaplan,
  %``Flavor at SSC energies: A New mechanism for dynamically generated fermion masses,''
  Nucl.\ Phys.\ B {\bf 365}, 259 (1991).
  doi:10.1016/S0550-3213(05)80021-5
  %%CITATION = doi:10.1016/S0550-3213(05)80021-5;%%
  %245 citations counted in INSPIRE as of 11 Jul 2016
  
  %\cite{Pica:2016rmv}
\bibitem{Pica:2016rmv} 
  C.~Pica and F.~Sannino,
  %``Anomalous Dimensions of Conformal Baryons,''
  arXiv:1604.02572 [hep-ph].
  %%CITATION = ARXIV:1604.02572;%%
  %3 citations counted in INSPIRE as of 11 Jul 2016
  
  %\cite{Sannino:2016sfx}
\bibitem{Sannino:2016sfx} 
  F.~Sannino, A.~Strumia, A.~Tesi and E.~Vigiani,
  %``Fundamental partial compositeness,''
  arXiv:1607.01659 [hep-ph].
  %%CITATION = ARXIV:1607.01659;%%
  
  %\cite{Holdom:1981rm}
\bibitem{Holdom:1981rm} 
  B.~Holdom,
  %``Raising the Sideways Scale,''
  Phys.\ Rev.\ D {\bf 24}, 1441 (1981).
  %%CITATION = PHRVA,D24,1441;%%
  %558 citations counted in INSPIRE as of 02 Nov 2015
  
  
 
  %\cite{Fukano:2010yv}
\bibitem{Fukano:2010yv}
  H.~S.~Fukano and F.~Sannino,
  %``Conformal Window of Gauge Theories with Four-Fermion Interactions and Ideal Walking,''
  Phys.\ Rev.\ D {\bf 82} (2010) 035021
  [arXiv:1005.3340 [hep-ph]].
  %%CITATION = ARXIV:1005.3340;%%
  %50 citations counted in INSPIRE as of 28 Aug 2015
  
    %\cite{Yamawaki:1996vr}
\bibitem{Yamawaki:1996vr}
  K.~Yamawaki,
  %``Dynamical symmetry breaking with large anomalous dimension,''
  hep-ph/9603293.
  %%CITATION = HEP-PH/9603293;%%
  %44 citations counted in INSPIRE as of 28 Aug 2015
  
  
  %\cite{Bitar:1993cs}
\bibitem{Bitar:1993cs}
  K.~M.~Bitar and P.~M.~Vranas,
  %``The Nambu-Jona-Lasinio model of QCD on the lattice,''
  Phys.\ Lett.\ B {\bf 327} (1994) 101
  [hep-lat/9310008].
  %%CITATION = HEP-LAT/9310008;%%
  %9 citations counted in INSPIRE as of 09 sept. 2015

%\cite{Bitar:1993xi}
\bibitem{Bitar:1993xi}
  K.~M.~Bitar and P.~M.~Vranas,
  %``A Study of the Nambu-Jona-Lasinio model on the lattice,''
  Phys.\ Rev.\ D {\bf 50} (1994) 3406
  [hep-lat/9310027].
  %%CITATION = HEP-LAT/9310027;%%
  %15 citations counted in INSPIRE as of 28 Aug 2015
  
  %\cite{Aoki:1993vs}
\bibitem{Aoki:1993vs}
  S.~Aoki, S.~Boettcher and A.~Gocksch,
  %``Spontaneous breaking of flavor symmetry and parity in the Nambu-Jona-Lasinio model with Wilson fermions,''
  Phys.\ Lett.\ B {\bf 331} (1994) 157
  [hep-lat/9312084].
  %%CITATION = HEP-LAT/9312084;%%
  %17 citations counted in INSPIRE as of 28 Aug 2015
  
  %\cite{Hasenfratz:1991it}
\bibitem{Hasenfratz:1991it}
  A.~Hasenfratz, P.~Hasenfratz, K.~Jansen, J.~Kuti and Y.~Shen,
  %``The Equivalence of the top quark condensate and the elementary Higgs field,''
  Nucl.\ Phys.\ B {\bf 365} (1991) 79.
  %%CITATION = NUPHA,B365,79;%%
  %169 citations counted in INSPIRE as of 14 sept. 2015
  
%\cite{AliKhan:1993dx}
\bibitem{AliKhan:1993dx}
  A.~Ali Khan, M.~Gockeler, R.~Horsley, P.~E.~L.~Rakow, G.~Schierholz and H.~Stuben,
  %``The Nambu-Jona-Lasinio model with staggered fermions,''
  Nucl.\ Phys.\ Proc.\ Suppl.\  {\bf 34} (1994) 655
  [hep-lat/9401003].
  %%CITATION = HEP-LAT/9401003;%%
  %2 citations counted in INSPIRE as of 28 Aug 2015
  
  %\cite{Kogut:1998rg}
\bibitem{Kogut:1998rg}
  J.~B.~Kogut, J.~F.~Lagae and D.~K.~Sinclair,
  %``Thermodynamics of lattice QCD with chiral four fermion interactions,''
  Phys.\ Rev.\ D {\bf 58} (1998) 034504
  [hep-lat/9801019].
  %%CITATION = HEP-LAT/9801019;%%
  %20 citations counted in INSPIRE as of 09 sept. 2015
  
  %\cite{Sinclair:1998ji}
\bibitem{Sinclair:1998ji}
  D.~K.~Sinclair, J.~B.~Kogut and J.~F.~Lagae,
  %``Thermodynamics of lattice QCD with massless quarks and chiral four fermion interactions,''
  Nucl.\ Phys.\ Proc.\ Suppl.\  {\bf 73} (1999) 471
  [hep-lat/9809052].
  %%CITATION = HEP-LAT/9809052;%%
  %2 citations counted in INSPIRE as of 28 Aug 2015
  
  %\cite{Catterall:2011ab}
\bibitem{Catterall:2011ab}
  S.~Catterall, R.~Galvez, J.~Hubisz, D.~Mehta and A.~Veernala,
  %``Non-abelian gauged NJL models on the lattice,''
  Phys.\ Rev.\ D {\bf 86} (2012) 034502
  [arXiv:1112.1855 [hep-lat]].
  %%CITATION = ARXIV:1112.1855;%%
  %8 citations counted in INSPIRE as of 28 Aug 2015


  %\cite{Bochicchio:1985xa}
\bibitem{Bochicchio:1985xa}
  M.~Bochicchio, L.~Maiani, G.~Martinelli, G.~C.~Rossi and M.~Testa,
  %``Chiral Symmetry on the Lattice with Wilson Fermions,''
  Nucl.\ Phys.\ B {\bf 262} (1985) 331.
  %%CITATION = NUPHA,B262,331;%%
  %424 citations counted in INSPIRE as of 08 sept. 2015


%\cite{Izubuchi:1998hy}
\bibitem{Izubuchi:1998hy}
  T.~Izubuchi, J.~Noaki and A.~Ukawa,
  %``Two-dimensional lattice Gross-Neveu model with Wilson fermion action at finite temperature and chemical potential,''
  Phys.\ Rev.\ D {\bf 58} (1998) 114507
  [hep-lat/9805019].
  %%CITATION = HEP-LAT/9805019;%%
  %15 citations counted in INSPIRE as of 14 sept. 2015


 
\end{thebibliography}
\end{document}